\documentclass[runningheads]{llncs}
\usepackage{graphicx}
\usepackage{amsfonts} 
\usepackage[binary-units]{siunitx}
\DeclareSIUnit{\Bit}{Bit}
\DeclareSIUnit{\Byte}{Byte}
\begin{document}
%
%\title{Contribution Title\thanks{Supported by organization x.}}
\title{A study on the use of perceptual hashing to detect manipulation of embedded messages in images}
\titlerunning{Perceptual hashing to detect manipulation of embedded images}
% If the paper title is too long for the running head, you can set
% an abbreviated paper title here
%
%\author{First Author\inst{1}\orcidID{0000-1111-2222-3333} \and
%Second Author\inst{2,3}\orcidID{1111-2222-3333-4444} \and
%Third Author\inst{3}\orcidID{2222--3333-4444-5555}}
\author{Sven-Jannik Wöhnert\inst{1}\orcidID{0000-0002-5850-9147} \and
Kai Hendrik Wöhnert\inst{1}\orcidID{0000-0002-1346-8126} \and
Eldar Almamedov\inst{1}\and
Carsten Frank\inst{1}\and
Volker Skwarek\inst{1}\orcidID{0000-0001-5065-1029}}
\authorrunning{S-J. Woehnert et al.}
% First names are abbreviated in the running head.
% If there are more than two authors, 'et al.' is used.
%

\institute{Digital Business Processes Research and Transfer Centre, Hamburg University of Applied Sciences, Ulmenliet 20, 21033 Hamburg, Germany
\email{Sven-Jannik.Woehnert@haw-hamburg.de}
}
\maketitle              % typeset the header of the contribution
\begin{abstract}
%Typically, metadata of images are  stored in a specific data segment of the image file, e.g. EXIF tags. However, for security reasons data can also be merged with images so that changes in that message or modifications of the image can be easily detected. In those cases, the embedding of the message follows the goal to invisibly embed as much information as possible and to make the embedding as robust as possible. Ideally, the message should even survive at least some form of transformation or compression.

%This work searches for embedding principles  which allow to distinguish  between unintended changes by lossy image compression and the intended, malicious manipulation of the specific message that is embedded in the image based on the change of its perceptual or robust hash. This problem is not trivial, since both processes, due to the required imperceptibility, can be implemented as a small modulation of the pixel values of the image. In this paper, different embedding and compression algorithms are introduced and compared for their usability with respect to this question. 
%The study shows that embedding a message via integer wavelet transform and compression with Karhunen-Loeve-transform yields the best results. However, it was not possible to distinguish between manipulation and compression in all cases.

% Kai's Vorschlag:
Typically, metadata of images are stored in a specific data segment of the image file. However, to securely detect changes, data can also be embedded within images. This follows the goal to invisibly and robustly embed as much information as possible to, ideally, even survive compression.

This work searches for embedding principles which allow to distinguish between unintended changes by lossy image compression and malicious manipulation of the embedded message based on the change of its perceptual or robust hash. Different embedding and compression algorithms are compared.

The study shows that embedding a message via integer wavelet transform and compression with Karhunen-Loeve-transform yields the best results. However, it was not possible to distinguish between manipulation and compression in all cases.

\end{abstract}

\section{Introduction}

%\copyrightspace

Associating meta information with images is common since the early days of the photography. This ranges from date, time and place where an image was taken up to  semantics such as  "Grandma with Peter at Christmas 1992" written on the back of the image. Nowadays, most images are taken with a digital device that automatically adds messages to the technical meta information. The message is commonly embedded in a dedicated part of an image file, for example to the "exchangeable image file format information" (EXIF) in the case of jpeg files. However, from the security perspective, the message must be more closely connected with the picture as EXIF can easily be modified and replaced. Any intentional modification of the picture or the metadata must be easily detectable. This security requirement excludes for example procedural modifications of an image by lossy compression, which aims to reduce the file size without visible changes. A proof that information in an image is untampered is important for all use cases where image and data integrity play an important role for documentation purposes in legal context. 

Message embedding needs to balance three factors: robustness, data volume and visibility \cite{Shete.2016}. Any significant change in the latter disturbs the perception of the image. This change in perception is commonly measured by the peak signal-to-noise ratio (PSNR) \cite{korhonen2012peak}.
Robustness and data volume of the embedded messages are more or less anticorrelated, so increasing the robustness of the embedded message also increases its size and, therefore, less content can be stored in the same data volume. Robustness is enhanced by error correction coding and refers to the ability to extract the message despite memory errors. A comparison of methods for data extraction methods is not part of this work.

To secure an image, different approaches have already been taken in recent research:

An early approach to image or video  security was to generate  cryptographic signatures and store the private key inside the camera \cite{friedman_trustworthy_1993}.

A similar approach is described by Danko et. al. \cite{danko_assuring_2019} where the hash value of a frame is sent to the server of a local authority, which is considered trustworthy.

In \cite{gennaro_how_1997} the authenticity of a video stream is confirmed by a signature of the first data block. A block also contains the hash of a successor. It is therefore necessary to know the last block before sending the first block.

To ensure scalability in authenticating videos, \cite{atrey_scalable_2007} proposes to identify key frames around which other frames vary little in time and calculate the standard derivation of the other frames from the key frame. Multiple key frames are signed together. 

In \cite{Wohnert.2020} the ideas were extended to a new approach of linking the frames similar to a blockchain, which aims to secure video streams for integrity and authenticity during recording. Wöhnert et. al. suggests, that three requirements have to be fulfilled: To be able to extract embedded information without external information (Autarky), be able to provide proof of integrity for subsamples of videostream (Modularity) as well as compressed videostreams (Robustness).

With robustness, the freedom to allow small deviations comes at a price: It allows small manipulations. As our proposed principle embeds a robust hash of a frame in  upcoming frames to secure the video sequence, another question arises: can the extracted hash value be trusted? Is it possible that an attacker has manipulated the embedded hash value in his favour to insert fake sequences? 

Three research questions are to be answered in this work:
\begin{enumerate}
    \item[RQ1] Is there an embedding algorithm where even small changes in the embedded message can be detected?
    \item[RQ2] Can the intensity of the change be quantitatively estimated?
    \item[RQ3] Is it possible to distinguish between allowed changes like compression and manipulations of the embedded message based on the hash value?
\end{enumerate}

In \cite{Wang.2021}, RQ3 has, in its own way, already been answered. Wang et al. used pre-compression to find pixels, which behave inert to the compression. These pixels are used to embed the message before compression. However, this is steganographic embedding, since without the knowledge of the exact pixel positions the message cannot be extracted. But this method cannot be used here due to the required property of autarky.

To answer the research questions above, we first introduce in chapter 2 the Hamming distance, which makes the difference between two robust hashes measurable. Then, various basic embedding algorithms and compression algorithms are introduced.
In chapter 3, the experiment is described in detail. In chapter 4, the evaluation follows and in chapter 5, the findings are summarized. 
\section{Related Work}
 
In this chapter, the robust hash, which is also called perceptual hash, will be explained.
To analyze whether and how much a robust hash has changed, the Hamming distance and the peak signal-noise-ratio (PSNR) are introduced.
Furthermore, different embedding and compression algorithms which will be used in the study are described. Those algorithms will be used to find the best algorithms to answer RQ1.

\subsection{Robust Hashing}

To authenticate a data set, typically cryptographic hash values are used. However, if a single data point changes, the associated cryptographic hash value changes completely/chaotically. Therefore, the robust hash is used instead. To make a hash robust against small allowed changes, robust hash is used. In general, the more the image changes, the more the robust hash changes. \cite{gharde2018robust}. For this paper, the block hash introduced by  \cite{yang2006block} is used. The change of the hash is introduced by a relative intensity change in a sub block of the image.

\subsection{Image Analysis}
The peak signal-to-noise ratio (PSNR) is a simple measure for the quality of an image. A low PSNR value indicates high distortion. A human cannot detect any distortion with grayscale images above 36 dB \cite{kumar2013comparative}. 

The PSNR value is described in equation \ref{eq:psnr} with x and y representing the dimensions of the image and p and p' respectively represent the pixel values of the image before and after edit.

\begin{equation}
    PSNR = 10 \cdot log_{10}\left(\frac{x \cdot y \cdot 255^2}{\sum\left(p-p'\right)^2} \right)
    \label{eq:psnr}
\end{equation}

The Hamming distance ($H_d$) is used to compare the similarity of two bit strings of the same length by counting differences at bit positions. The distance is an indicator of noise or other changes in an image. The smaller the Hamming distance, the higher the probability that the image is perceptually the same \cite{jegou2008hamming}. The Hamming distance is the percentage of bit-switches in robust hash between original image and edited image.

\subsection{Embedding Algorithms}

Embedding a message in the least significant n bit(s) of pixel values is a technique with high data volume and a balancing between robustness and visibility \cite{ElhamGhasemi.2012}. This method to embed messages can be used for both message-in-image and image-in-image \cite{bender_techniques_1996}, like e.g. QR codes.

QR is a very popular technology for the graphical representation of text or binary message \cite{hajduk_image_2016}. It uses Reed-Solomon-ECC and orientation bits and this has robustness against image rotation and bit errors.

Alternatively, the information can be embedded into any other domain, e.g. the frequency domain. The most popular frequency domain algorithm is the discrete cosine transform (DCT). 
Although it is mostly used for compression, it is also suitable for embedding message. Embedding in frequency space, especially in lower frequencies, provides high robustness and allows extraction of the message even after significant image changes. As coefficients of a DCT are not integer, the embedding strength is not controllable and depends on the pixel value. Therefore, the M-ary quantisation index modulation (QIM) from \cite{Chen.2001} is used. Here, a coefficient X is modulated up or down by the quantization quantity $q_s$ depending on the bit value $m$ of the message, see equation \ref{eq_qim} from \cite{Moulin.2005}.

\begin{equation}
    X' = int\left(\ \frac{X}{q_s} \right) q_s + \left(-1\right)^{m+1} \cdot \frac{q_s}{4}
    \label{eq_qim}
\end{equation}

The discrete wavelet transform (DWT) is another frequency domain algorithm. In contrast to DCT, in DWT data is convoluted with a wavelet and is popular for data embedding \cite{ElhamGhasemi.2012}. As with DCT, the core of the data is in the low frequencies, for DWT in the LL band. Embedding has again to be combined with QIM.

The integer wavelet transform (IWT) is a mixed form of a spacial and frequency domain transform \cite{ElhamGhasemi.2012}. Equations \ref{eq:iwt_low} and \ref{eq:iwt_high} describe the formula for the haar wavelet. $X_i$ is the i-th row or column of the image. g and h are formed out of two adjacent rows or columns. The floor-function rounds to the lower or equal integer. Like LSB, IWT can embed a comparably large amount of data while being more robust compared to DCT.

\begin{equation}
h_j = \mathrm{floor}\left(\frac{X_i + X_{i+1}}{2}\right)
\label{eq:iwt_low}
\end{equation}

\begin{equation}
g_j = X_i - X_{i+1}
\label{eq:iwt_high}
\end{equation}

\subsection{Compression Algorithms}
    
One goal of image compression is the reduction of resources for storing, sending and processing images. Compression algorithms can be categorized into lossless, near-lossless, and lossy compression. In this article, we will focus on lossy compression in which the source image is not fully recoverable but higher compression ratios compared to lossless and near-lossless compression methods can be achieved. E.g., lossy methods transform images into the discrete spectra and the coefficients are quantized and truncated so that information is lost. Lossy methods are therefore problematic for message embedding as compression algorithms may destroy parts of the message. Below, different combinations of embedding methods and compression algorithms are discussed.

In DWT and DCT, the image is transformed into the frequency domain. DWT has been optimized since its introduction by Mallat in 1987 \cite{mallat1987theory} and is widely used due to its feature support for image compression. DWT is the basis of JPEG2000 \cite{acharya2006survey}.

DCT is a much faster algorithm than DWT, when implemented on application-specific integrated circuits, and is widely used for speech and HD TV \cite{rao2014discrete}. The underlying methods to achieve a DCT can vary from sparse matrix factorization, fast Fourier transformation, or other discrete transforms \cite{rao2014discrete}.

Quadtree and spline interpolation are methods used in image compression working in the spatial domain. Quadtree images are divided into blocks and stored in a hierarchical data structure. Each block is either a leaf with no further subdivision or has four sub blocks describing the image in more detail. Depending on the resolution level, the number of hierarchy levels, also called depth, are chosen \cite{samet1984quadtree}.

Spline interpolation is a method used to smooth an image after reducing its pixels by interpolating between discrete points \cite{truong2000image}. It can reduce visual distortion after compression, it is not intended to reconstruct the original image.

The Karhunen–Loève transform (KLT) is a frequency based algorithm similar to DCT. But instead of a uniform basis like cosines with different periodicity, the transform matrix consists of the eigenvectors of the image covariance matrix \cite{Hua.1998}. This allows a much higher compression quality but also requires higher computational performance.
\section{Method}
This chapter introduces measures for the experiments concerning the sensitivity between intended message modifications by intended compression and unintended manipulation. Measures to quantify the embedding process are described in section \ref{sec_embedding_algos} and measures for compression tolerance are described in section \ref{sec_compression_algos}.
All experiments were implemented with python 3.8. The algorithms were implemented using python packages (scipy.fftpack for DCT, pywt for DWT, reedsolo for Reed-Solomon-ECC, qrcode for QR code, scipy.interpolate.interp2d for SPLINE), based on publications (QIM \cite{Moulin.2005}, IWT \cite{V..2019}) or adapted from already implemented projects (KLT \cite{SonuDileep.}, QUADTREE \cite{alpharaoh.}). The common test images Lenna, Baboon and Peppers as PNG in RGB mode with resolution 512x512 were used.

\subsection{Embedding Algorithms}
\label{sec_embedding_algos}
A selection of four algorithms was made, which allow a bitwise embedding and extraction of a message. In order to compare the algorithms, the embedding strength of each algorithm is chosen so that the PSNR equals to $36\pm \SI{0.5}{\dB}$. For grayscale images, 36dB is the perception threshold for the human visual system \cite{R.O.ElSafy.2009}. As no criterion exists for colour images, $\SI {36}{\dB}$ is also used for the perception threshold for the color images.

The message was transformed into a QR code and embedded into the least significant bit (LSB) of the picture. The QR code facilitates the retrieval of the message and adds additional redundancy by error correction coding (ECC). As a second spatial domain method, IWT was used. To increase robustness, the message was embedded only in the LL band. Subsequently, an inverse IWT was performed. Experiments with a LSB in range of [1,7] have been executed. Using the last 3 bits, the PSNR has been shown to be within the required range of values for both QR code embedding and IWT.
The following steps were performed for both algorithms:

\begin{enumerate}
    \item Divide the messages into 3 equal parts
    \item Determining the block size $b_{qr} = n^i * b_{hash} \land b_{qr}\left(n^i\right) * qr_{size} <= dim\left[image\right]$. $b_{qr}$ is the block size from the QR code, $b_{hash}$ is the block size from the block hash. With $n \in \mathbb{N}$ and $i \in \left[-1,1\right]$ block sizes are multiples of each other. $n^i$ is maximized.
    \item IWT-only: Extract LL-band as carrier image.
    \item To prevent the position and orientation bits from overlapping, the 3 QR codes are embedded in different corners of the 3 color channels in LSB-3.
\end{enumerate}
For the QR code, the chosen amount of pixels which represent one bit increases the sensitivity of the Hamming distance. If the QR code is manipulated by a third party, a whole group of pixels has to be altered to change a single bit. If the group of pixels is within a block of the block hash, then the mean of the pixel values in the block will change significantly. This maximizes the chance that this change will exceed the threshold of a block and thus increase the hamming distance.

To embed a message in the coefficient of a frequency space transform, QIM is used as described above. This allows embedding and extraction of a bit in a floating point number with a defined embedding strength. DCT with a block size of 8x8 and DWT with the Haar wavelet are used as representatives for the frequency space transformation. Similar to IWT, the embedding is done in the 4x4 block of the low frequencies and in the LL band, respectively. From simulations with quantization sizes from [10,80] it has been found that for DCT $q_s=23$ and for DWT $q_s=21$ should be chosen.

\begin{figure}[htb]
    \centering
    \includegraphics[width=0.9\textwidth]{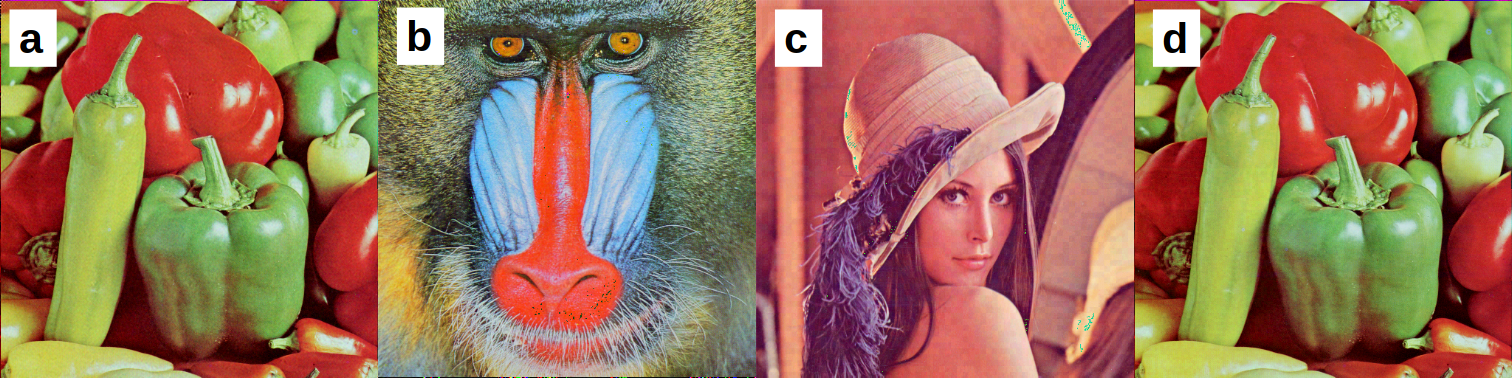}
    \caption{Image embedding adjusted for PSNR$\approx \SI{36} {\dB}$. a) DCT embedding in Peppers with $q_s = 23$ and PSNR$ = \SI {35.92}{\dB}$. In central areas a green pixel noise is visible. b) DWT embedding in Baboon with $q_s = 21$ and PSNR$ = \SI{35.84}{\dB}$. c) IWT embedding in Lenna with LSB=3 and PSNR$=\SI{35.68}{\dB}$. Noise in large areas such as the small yellow diagonal bar in the top right corner is visible. d) QR embedding in Peppers with LSB=3 and PSNR$=\SI{35.80}{\dB}$.}
    \label{fig:max_psnr}
\end{figure}

The experiment was conducted as follows:
\begin{itemize}
    \item 10 Message elements containing a $\SI{16} {\Byte}$ hash and a $\SI{7}{\Byte}$ timestamp are randomly generated
    \item The message is embedded in the test images.
    \item PSNR and Hamming Distance between source image and embedded message is determined.
    \item Exchange of message elements in range [1,10]. First, replace one out of 10 elements, then replace two elements etc. Determination of PSNR and Hamming distance between original message and manipulated message.
    \item Each exchange was executed 20 times. The new message element is calculated from random seed. Mean value and standard derivation were calculated.
\end{itemize}

The goal of this experiment is to determine a dependence of the Hamming distance on the strength of the manipulation. The strength of manipulation extends from 0 (no message element manipulated) to 1 (all 10 message elements manipulated). In addition, the embedding algorithms will be compared. This tests the hypothesis that the Hamming distance does not correlate with the PSNR value.

\subsection{Compression Algorithms}
\label{sec_compression_algos}
In this chapter different lossy compression algorithms will be examined in terms of PSNR and Hamming distance. 
\begin{itemize}
\item frequency domain: DCT, DWT (with Haar wavelet) and KLT. KLT is a computationally expensive and (therefore) rarely used algorithm, however, it excels in the ratio of image quality to memory savings.
\item spatial domain: Quadtree and Spline. Quadtree, like the robust Blockhash, is based on blocks, where during compression each pixel of a block takes the value of the block mean. This may prove to be an advantage with respect to spatial domain embedding algorithms. The fact that an extraction of the embedded message is virtually impossible with this compression is to be neglected here. The spline algorithm provides smoothing of the image while preserving selected rows and columns.
\end{itemize}
A selection of compression levels is chosen for each algorithm. The compression level is the ratio of the storage requirement of the compressed image and the original image. For DCT, from 70\% down to 2\% was chosen. For DWT, a triple LL band extraction was chosen (25\%, 6.25\%, 1.5\%).  For KLT, a block size of 8x8 is chosen, resulting in 64 eigenvectors used in the transformation. The reduction of eigenvectors was chosen so that the compression level is approximately the same as DCT. For Quadtree, the maximum depth of the leaves in the tree was chosen between 3 and 8. For Spline, the rows [2,7] were conserved. Then, for each algorithm, test pattern and compression level, the Hamming distance was determined.
Since the compression algorithms are deterministic, each step is performed once.
\section{Experimental Results}

\subsection{Embedding Results}

\begin{figure}[htb]
    \centering
    \includegraphics[width=0.9\textwidth]{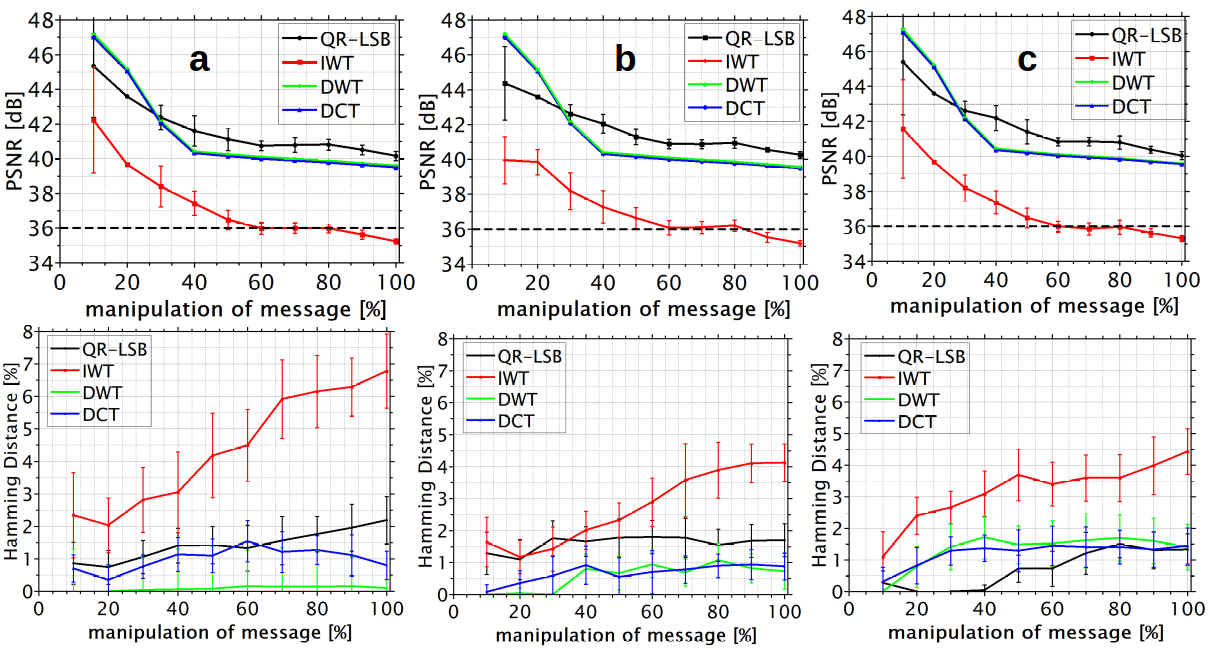}
    \caption{PSNR value and Hamming distance for embedding in the test images a) Baboon, b) Lenna and c) Peppers. The X-axis shows the number of manipulated messages as a percentage. A total of 10 messages are embedded. The Y-axis shows the PSNR value in decibels with a marker at $\SI {36}{\dB}$ and the Hamming distance in percent.}
    \label{fig:embedding}
\end{figure}

The results in figure \ref{fig:embedding} show that IWT is the most sensitive to manipulation.
This is shown for PSNR with values even below $\SI {36}{\dB}$ as well as for the Hamming distance with up to 7\% of switched bits. While the PSNR value is mostly the same for all test images, the maximum mean Hamming distance varies between 4\% (Lenna) and 7\% (Baboon).
For frequency domain methods DCT and DWT, the change of the PSNR value is the same. The error bars only occasionally deviate from zero, which means the noise pattern is the same regardless the replaced data of the message.
The Hamming distance criterion results show that the frequency domain methods react sensitive to the choice of the test image. The mean Hamming distance is partially close to zero for the test images Baboon and Lenna, therefore, an identification of a manipulation would not be possible.
Overall, embedding a QR code has the best PSNR value and is second best using the Hamming distance criterion. The latter can be attributed to the choice of block size of the QR image. However, in the case of the test image Peppers, no identification of the manipulation is possible up to 40\%.
As a result, frequency domain based methods are not suitable for identifying tampering. IWT performs best except for the fact that a large manipulation even falls below the threshold of $\SI{36}{\dB}$, causing a similar amount of noise as the original embedding. The manipulation of the QR code causes little noise, but also has a lower Hamming distance than IWT. Despite the too low Hamming distance for a homogeneous image like Peppers, the algorithm can still be considered as a good method for embedding messages regarding research question RQ1.

\subsection{Compression Results}

\begin{figure}[htb]
    \centering
    \includegraphics[width=0.9\textwidth]{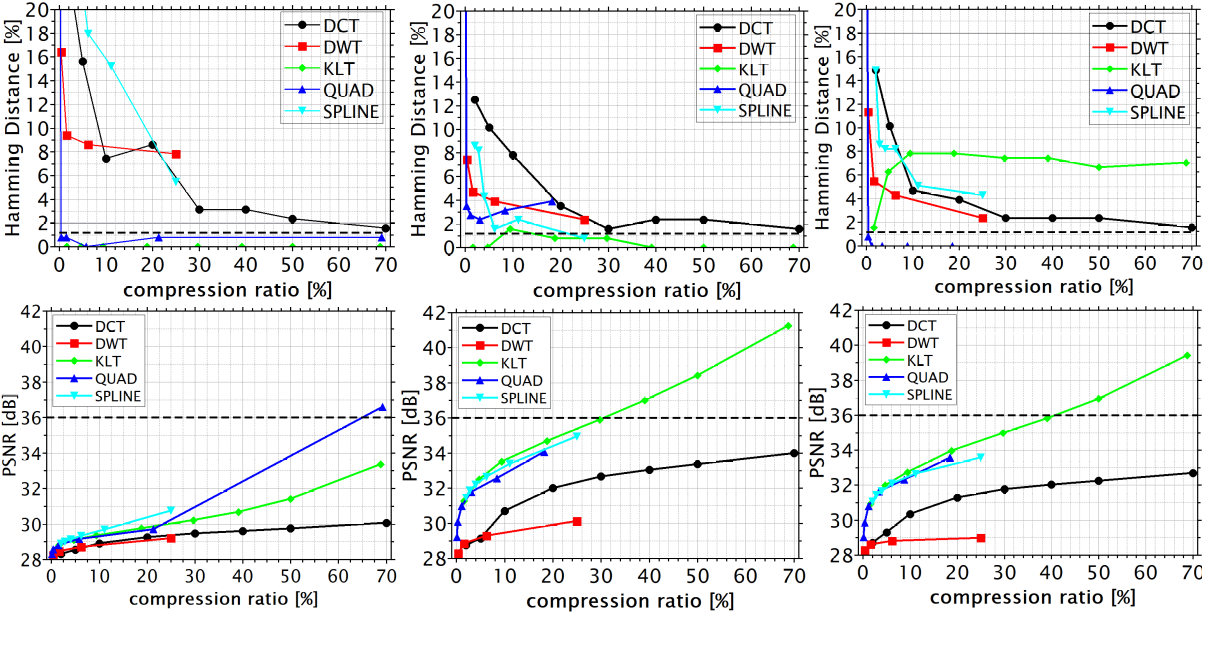}
    \caption{PSNR value and Hamming distance for the compression of test images a) Baboon, b) Lenna and c) Peppers. The X-axis shows the compression ratio measured in disk space compared to the original image. The Y-axis shows the PSNR value in decibels with a marker at 36dB and the Hamming distance with a marker at the lowest mean value for IWT embedding for comparison.}
    \label{fig:compression}
\end{figure}

Compressing an image is a frequently used intended image manipulation that should still be possible without changing the message. Compression has an effect on the same criteria as all other manipulations.
Our experiments show, that the more heterogeneous the image, the smaller the PSNR value (see figure \ref{fig:compression}). For Baboon, the PSNR value is less than $\SI{30}{\dB}$ for all algorithms and all compression ratios and therefore the manipulations are visible, while for Lenna and Peppers, the values are well above $\SI {30}{\dB}$ at a compression of 20\% or higher. The Hamming distance results show, that for Baboon both KLT and Quadtree are below the minimum Hamming distance of an IWT manipulation. In the case of KLT, due to the high quality of the compression, the Hamming distance is close to zero which enables a distinction between manipulation of IWT embedding and KLT compression. Quadtree also achieves good results for a depth of at least three iterations. In this case the compression block size of the Quadtree is less than or equal to the block size of the hash. Due to the nature of the Quadtree method, the compressed value of a Quadtree block is equal to the mean value of the original pixels of the block.
For Lenna, Quadtree is above the defined threshold for all depths, while KLT has an outlier at 10\% message manipulation. The results of the image Peppers are partially unexpected as the Hamming distance is approximated constant at 8\% for low compression between 70\% and 10\% but decreases for higher compressions. Quadtree is again below the threshold for depths from 3 iterations.
KLT, just like other frequency space based compression algorithms, does not cope well with partially very homogeneous images like Peppers as seen in the embedding in Figure \ref{fig:max_psnr}.
\section{Conclusion}
The embedding experiment has clearly shown that IWT is the most suitable algorithm for embedding messages into an image. With respect to research question RQ1, it can be said that with IWT it is possible to identify manipulations. However, it cannot be guaranteed that the Hamming distance is greater than zero in every case. This means that a potential attacker can statistically find a frame to manipulate the message without trace as an entry point to the hashed chain of video frames. A possible solution for this problem could be to form a cryptographic hash of the message, which is also embedded in the next frame. However, this approach has yet to be tested. In regard of RQ2, IWT is also the only embedding algorithm to show a causality between percent of message manipulated and Hamming distance.

The compression experiment has shown that KLT and Quadtree compression can be distinguished from message manipulation, disregarding homogeneous images like Peppers and a compression to less than 10\%. The answer to RQ3 is: Although the compression strength can be specified, it is not possible to prevent a homogeneous image from being used. Therefore, when the robust hash is changed, it is not possible to decide whether this was done by compression or manipulation of the message.

To solve the problem, it is useful to look at other evaluation parameters besides block hash. In addition, only the most common embedding algorithms were tested, perhaps there is a variant among those not yet tested that fulfills the research questions posed. For example Singular Value decomposition is a promising candidate \cite{goos_svd-based_2001}.
When all options are exhausted, any kind of lossy compression should be considered as a tampering attempt. Despite this restriction, the user is still left with compression using lossless compression algorithms. Another effect of this restriction is that extracting the message after compression is no longer problematic.

\section*{Acknowledgment}
This research was performed within the project TrustedCam and supported by the German Ministry of Education and Research (BMBF Grant No. 13FH214PX8 .msg).

The authors of these publications have committed themselves to the guidelines of good scientific practice of the German Research Foundation. To ensure the quality of the publication, all Data and Code are publicly accessible via DOI
10.17605/OSF.IO/BJHM4.

\bibliographystyle{ieeetr}
%\bibliography{bibliography}

\end{document}